# Magneto-thermoelectric figure of merit of Co/Cu multilayers

X.K. Hu,[1,*] P. Krzysteczko[1], N. Liebing[1], S. Serrano-Guisan[2], K. Rott[3], G. Reiss[3], J. Kimling[4], T. Böhnert[4], K. Nielsch[4], and H.W. Schumacher[1]

[1]Physikalisch-Technische Bundesanstalt, Bundesallee 100, D-38116 Braunschweig, Germany

[2]International Iberian Nanotechnology Laboratory, Av. Mestre José Veiga, 4715-330 Braga, Portugal.

[3]Fakultät für Physik, Universität Bielefeld, Postfach 100131, D-33501 Bielefeld, Germany

[4]Institut für Angewandte Physik, Universität Hamburg, Jungiusstraße 11 B, 20355 Hamburg Germany

*Corresponding author:

Tel: +49(0)531 592 1412

E-mail: xiukun.hu@ptb.de

**ABSTRACT**

The switching of the magnetization configurations of giant magnetoresistance multilayer stacks not only changes the electric and thermal conductivities, but also the thermopower. We study the magnetotransport and the magneto-thermoelectric properties of Co/Cu multilayer devices in a lateral thermal gradient. We derive values of the Seebeck coefficient, the thermoelectric figure of merit, and the thermoelectric power factor. The Seebeck coefficient reaches values up to -18 µV/K at room temperature and shows a magnetic field dependence up to 28.6 % upon spin reversal. In combination with thermal conductivity data of the same Co/Cu stack, we find a spin dependence of the thermoelectric figure of merit of up to 65 %. Furthermore, a spin dependence of the power factor of up to 110 % is derived.

The discovery of the giant magnetoresistance (GMR) effect in multilayers consisting of alternating ferromagnetic and non magnetic metal layers in 1988 [1] triggered the development of spintronics [2]. Extensive investigations have been performed on multilayers due to important applications in data storage and magnetic sensors [3]. Nowadays, the new field of spin-caloritronics [4, 5] addresses the combination of thermoelectricity and spintronics and has led to an increased interest in magneto-thermoelectric effects.

The thermoelectric figure of merit $ZT = \sigma S^2 T/\kappa$ and the power factor $PF = \sigma S^2$ are the key parameters to quantify and calculate the efficiency of thermoelectric devices. Here, $S$ is the Seebeck coefficient, $\sigma$ is the electric conductivity, $\kappa$ is the thermal conductivity, and $T$ is the temperature. In order to improve $ZT$ either $S$ should be increased or the Lorenz number $L = \kappa/\sigma T$ should be decreased as $ZT = S^2/L$. Measurements of the magneto-thermopower [6,7,8,9,10,11,12,13,14,15] and of the magneto-thermal conductivity [13-15,16] of GMR material systems have been carried out to evaluate the nature of spin-dependent scattering responsible for the GMR.

In metals, the thermopower follows the Mott formula [17] and is field-dependent when the spin asymmetry ratio (i.e. the ratio of two separate resistivities for each spin channel in a two-band GMR model) is energy-dependent [10,13]. The thermopower scales proportional to the electric conductance with the magnetic field as an implicit variable for many alloyed, multilayered, and granular systems [13,18,19,20]. This leads to the conclusion, that the energy derivative of the resistivity in Mott's equation is independent of the magnetic field. By changing the spin configuration of a multilayer, the thermopower should also be changed, as inferred by different theoretical models [11,13,21]. However, concerning the Lorenz number $L$, there are two controversial conclusions. Most of the works [6-13,10,14] confirmed that elastic scattering is responsible for GMR in magnetic multilayers, i.e. the Wiedemann-Franz law [22] is satisfied. The Lorenz number $L = \kappa/\sigma T$ is thus field-independent at constant temperature. Whereas Shi et al [13,15] and Yang et al [15] concluded that there are contributions from inelastic scattering to the GMR and that the $\kappa$ does not scale with $\sigma$. Here, $L$ changes by about 10 % between parallel and antiparallel spin configurations. Recently, Kimling et al. [16] revisited the concept of inelastic scattering on GMR in Co/Cu multilayers by employing the 3ω-method [23] to study the field dependent $\kappa$. They found that the Wiedemann-Franz law holds in the field-range of GMR even in the presence of weak inelastic scattering. However, despite of this comprehensive data on thermal conductivity and thermoelectricity of GMR materials, neither the magnetic field dependence of the thermoelectric figure of merit – the *magneto-ZT* – nor the field dependence of the power factor – the *magneto-PF* – [21] have been addressed so far.

In this work we combine studies of GMR and of the giant magneto-thermopower (GMTP) of Co/Cu multilayer devices with investigations of the magnetothermal conductivity of the same stacks [16]. The Seebeck coefficient $S = - V_{th}/\Delta T$ is derived from measurements of the thermopower $V_{th}$ in lateral thermal gradients $\Delta T$. $\Delta T$ is determined by local resistive thermometry in combination with finite element simulations of the temperature distribution. We derive Seebeck coefficients of $S_P = -18$ µV/K for parallel and $S_{AP} = -14$ µV/K for antiparallel spin configuration, a giant magneto-*ZT* ratio above 65 % and a giant magneto-*PF* of 110 %.

We study a sputter deposited Co/Cu multilayer stack of [Co(3 nm)/Cu(1 nm)]$_{39}$/Co(3 nm)/Ru(5 nm) on a fused silica substrate. The stack is patterned into 10 µm wide and 100 µm long microbars with 60 nm thick Pt contacts as shown in Fig. 1 (Co/Cu stripe: horizontal light bar in the center). The Pt contacts overlap the ends of the Co/Cu bar by 2 µm to ensure a good electric properties. The Pt contacts (1-8) are used to measure the electric properties, to generate lateral temperature gradients across the GMR stripe by Joule heating, and to measure the local temperature by Pt resistance thermometry. After wire bonding on a sample holder the sample was placed in an electromagnet allowing application of in-plane fields up to 900 mT.

The electric and thermoelectric characterization were carried out as follows: For GMR measurements, the four-probe resistance of Co/Cu multilayer was measured as function of the magnetic field at 100 $\mu$A bias current. For measurements of $V_{th}$, a temperature gradient between the ends of Co/Cu multilayer was generated by a DC current of 0.5 - 4 mA through one of the Pt lines used as a heater (e.g. between contacts 1 and 2). A nanovoltmeter was attached to two contacts across the Co/Cu bar (e.g. contacts 3 and 7) to detect $V_{th}$.

The resulting temperature gradient between the ends of the Co/Cu multilayer was quantified in the following way: During the experiments the resistance of the Pt lines between different contacts was used to probe the local temperature increase $\Delta T$ (starting at room temperature) on different positions using $R(T) = R_0(1+\alpha\Delta T)$. Here, $R(T)$ and $R_0$ are the resistances of the Pt lines when the heating current is switched on and off, respectively, and $\alpha$ is the temperature coefficient of resistance of platinum. $\alpha$ of the Pt lines was calibrated on a variable temperature probe station delivering $\alpha = 1.14\times10^{-3}$ K$^{-1}$ which is significantly smaller than the bulk value of $3.9\times10^{-3}$ K$^{-1}$. [24]. Based on the local temperatures derived from the resistance thermometry the temperature distribution in the sample was simulated using a 3D commercial finite element modeling tool [25].

Fig. 2 shows typical temperature data for a heating across contacts 1 and 2 and measuring temperature across contacts 7 and 8. The increase of the temperature of the Pt heater and thermometer with respect to room temperature is plotted as a function of the square of the heating current $I^2_{heater}$ and hence of the approximated applied heater power $P_{heater}$ (upper scale). The measured increase of the heater temperature $\Delta T_{heater}$ is given by the red dots and the measured increase of the thermometer $\Delta T_{7\text{-}8}$ by the red triangles. Both increase linearly with $P_{heater}$ as expected. The inset shows a result of the simulated temperature distribution for such heater configuration where a heater current of $I_{heater}$ = 4 mA is applied across contacts 1 and 2. The structure is sketched by the black lines and the local temperature is given by the color shade. From the model the temperature difference across the GMR structure $\Delta T_{GMR}$ can be derived. In the simulation $\kappa$ of 28.83 and 20.17 W/(m·K) of the Co/Cu multilayer for parallel (P) and antiparallel (AP) configurations, respectively, was used. Furthermore a heat conductivity of the substrate of 1.5 W/(m·K) was assumed as derived in [16].

From the simulations we obtain $\Delta T_{GMR}$ for both magnetic configurations as a function of $I^2_{heater}$. $\Delta T_{GMR}$ increases linearly as indicated by the blue dots (P) and blue open squares (AP), and reaches a maximum value of 4.6 K at $I_{heater}$ = 4 mA. Although the thermal conductivity of Co/Cu changes almost by 40 % upon magnetization switching, the temperature equilibration in the sample is largely dominated by the thick substrate. Therefore, the difference of the temperature gradients $\Delta T_{GMR}$ between parallel (P) and antiparallel (AP) magnetization configuration is only 40 mK for the maximum heating current and can be neglected. It is important to note that the simulated average temperature between contacts 7 and 8 (blue triangles) are in a very good agreement with the experimental data (red triangles) with less than 30 mK difference between simulation and experiment. This indicates that our simulated temperature distribution is a sound basis for the quantitative evaluation of *S* and *ZT*.

A typical GMR ratio *vs*. field curve at room temperature is shown in Fig. 3(a). The resistivity decreases with field and approaches saturation above 350 mT. The maximum GMR ratio $\Delta\rho/\rho_P = (\rho_{AP} - \rho_P)/\rho_P$ is about 27.2 %. As shown in the inset of Fig. 3(a), the GMR ratio decreases with increasing $\Delta T_{GMR}$ by a few % because of the increase of the average temperature [13,15,16]. To rule out that this variation changes the analysis of our data the magneto-thermoelectric measurements are carried out for low heater currents $I_{heater} \leq 2$ mA, i.e. for $\Delta T_{GMR} \leq 1.2$ K. Figure 3(b) shows $V_{th}$ as function of the magnetic field $\mu_0 H$ at room temperature with $I_{heater}$ = 2 mA applied. $V_{th}$ increases with field from about 15.9 μV (AP, *H* =

0) and saturates at 20.3 µV at about 350 mT (P). From the corresponding difference of $\Delta V_{th} = V_{th,P} - V_{th,AP} = 4.4$ µV we derive a GMTP ratio $\Delta V_{th}/V_{th,AP}$ of 27.7 %. The inset of 3(b) shows $V_{th}$ as a function of the electric conductance $\sigma = 1/\rho$. The plot compiles the experimental data of the main Figs. 3(a) and (b) and shows a linear dependence in agreement a two-band GMR model [9,10,13,14].

A similar behavior is also observed for different $\Delta T_{GMR}$. Fig. 4(a) plots $V_P$ and $V_{AP}$ vs. $\Delta T_{GMR}$ ($I_{heater} = 0.5 \ldots 2$ mA). $V_P$ and $V_{AP}$ both increase linearly with $\Delta T_{GMR}$. The negative slopes of linear fits to these data yields the Seebeck coefficients $S_P$ and $S_{AP}$ for parallel and antiparallel configurations of $S_P = -18$ µV/K and $S_{AP} = -14$ µV/K, yielding a GMTP ratio of 28.6%. The literature values of $S$ of bulk Co and Cu are $S_{Co} = -30.8$ µV/K [26] and $S_{Cu} = 1.84$ µV/K [27]. Analogue to the two-band model the Seebeck coefficient in in-plane direction of a multilayer can be estimated by weighting the Seebeck coefficient $S$ by the corresponding conductance of each layer yielding a reasonable agreement with the experimental data [28]. Further our data agrees with the values of Co/Cu multilayers reported previously [9,11-13].

Using these experimental values of $S_P$, and $S_{AP}$ along with the Lorenz number $L = 1.83 \times 10^{-8}$ V$^2$K$^{-2}$ [16] the field dependent $ZT$ can be derived. Fig. 4(b) shows $ZT$ derived both for P and AP configurations. $ZT$ only slightly varies with $\Delta T_{GMR}$ and yields values of $ZT_{AP} \approx 0.011$ (zero field) and of $ZT_P \approx 0.018$ (500 mT). From that we deduce a giant magneto-$ZT$ ratio of GMZT = $\Delta ZT/ZT_{AP}$ = $(ZT_P - ZT_{AP})/ZT_{AP}$ of about 65%. As shown in the figure $\Delta ZT/ZT_{AP}$ is constant above $\Delta T_{GMR} = 0.5$ K. We attribute the fluctuations of $ZT$ and of the GMZT at low heating currents to artifacts resulting from the residual noise dominating the low signal measured by the nanovoltmeter.

Furthermore, from the experimental values of $S_{AP}$, $S_P$, $\sigma_{AP}$, and $\sigma_P$ we deduce power factors of $PF = 1.77$ mW·m$^{-1}$·K$^2$ (P) and $0.84$ mW·m$^{-1}$·K$^2$ (AP). The magneto-$PF$ can be also written as $\frac{\Delta PF}{PF_{AP}} = \frac{\sigma_P S_P^2}{\sigma_{AP} S_{AP}^2} - 1 = \frac{\rho_{AP}}{\rho_P}\left(\frac{S_P}{S_{AP}}\right)^2 - 1 = \left(\frac{\Delta \rho}{\rho_P} + 1\right)\left(\frac{\Delta S}{S_{AP}} + 1\right)^2 - 1$. Hence we can deduce the *magneto-PF* ratio from the GMR and GMTP ratios given above yielding a giant magneto-*PF* (GMPF) up to 110 %.

It should be noted, that the GMTP in our sample is a bit higher than that in Refs. 11 and 12, but lower than that in Refs. 9 and 13. Such difference can be attributed to the different structures of the multilayers. For example, the samples [Co(15Å)/Cu(9Å)] in Refs. 12 and 29 differ only in the number of bilayers, but the sample in Ref. 29 exibits much higher GMR ratio and lower saturation field. Furthermore, with increasing the thickness of Cu layer, Co/Cu multilayers exhibit an oscillatory GMR, and GMR peaks occur in the antiferromagnetic half-periods [29]. As discussed previously, the GMTP effect is correlated with the GMR effect. The higher the GMR ratio, the higher GMTP can be expected [9,13]. For our sample, the thickness of Co layer is 3 nm, and thus higher than that in above references, which leads to a lower GMR due to shunting effects. A systematic investigation with different thickness of Cu layer could allow finding the best design of multilayers to achieve higher GMR and GMTP, and higher GMZT and GMPF ratios. Future investigations should also address the *magneto-ZT* of other promising spin-caloritronic materials such as magnetic tunnel junctions with large reported tunneling magneto-thermopower up to about 100 % [30,31]. Such materials could enable significantly higher absolute values of *ZT* as required for thermoelectric applications while keeping a high magneto-*ZT* to enable future magnetic field-controlled switchable thermoelectric machines.

In conclusion, we have investigated the thermoelectric properties of Co/Cu multilayer micro-stripes in lateral thermal gradients. The magneto-thermoelectric measurements yield Seebeck coefficients up to -18 µV/K with changes up to 28.6 % upon magnetization reversal. By combining this data with thermal conductivity data of the same stacks the thermoelectric figure of merit *ZT* has been derived. *ZT* yields a value of about 0.011 with a giant magneto-*ZT* change above 65%. Furthermore a thermoelectric power factor 0.84 mW·m$^{-1}$·K$^2$ with GMPF of 110% with a magnetic field change has been observed.

**ACKNOWLEDGMENTS**

We acknowledge discussions with G. Bauer on the importance of deriving *ZT* of spin-caloritronic materials. The research was funded by DFG SPP 1538 and by EMRP JRP EXL04 SpinCal. The EMRP is jointly funded by the EMRP participating countries within EURAMET and the EU.

**Figure Captions:**

**Fig. 1**: Scanning electron micrograph of a typical device and sketch of electric setup. Contacts numbered. Contacts 2 and 6 are used to applying 100 $\mu$A in GMR measurements; Contacts 1 and 2 are used for heating. $V_{th}$ is measured via contacts 3 and 7. In-plane magnetic field is applied along the multilayer bar.

**Fig. 2**: Temperature increase *vs.* $I^2_{heater}$. Measured increase (red symbols) of heater (dot) and thermometer $\Delta T_{7-8}$ (open triangle). Simulated temperatures $\Delta T_{GMR}$ (blue symbols) across GMR stripe for P (dot) and AP (open square) configurations and of thermometer $\Delta T_{7-8}$ (open triangle). Contour map shows the simulated temperature distribution at $I_{heater} = 4$ mA.

**Fig. 3**: Room temperature magnetic field dependences of the (a) GMR and (b) $V_{th}$. Insets: (a) temperature dependence of GMR; (b) $V_{th}$ *vs.* 1/$\rho$ at $I_{heater} = 2$ mA.

**Fig. 4**: (a) $V_{th}$ and (b) corresponding $ZT$ values *vs.* $\Delta T_{GMR}$ for P (filled square) and AP (dot) configuration. $S_{P,AP}$ is derived from fitting the slopes of (a). Open circles in (b) indicate a constant magneto-$ZT$ of about 65 %.

**Fig.1**

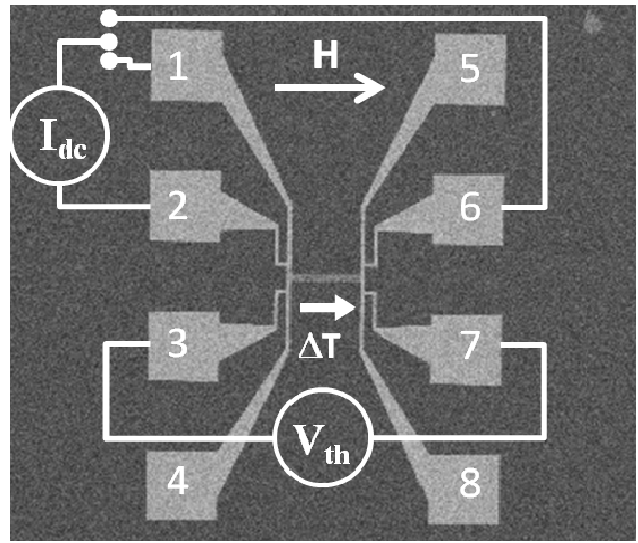

**Fig.2**

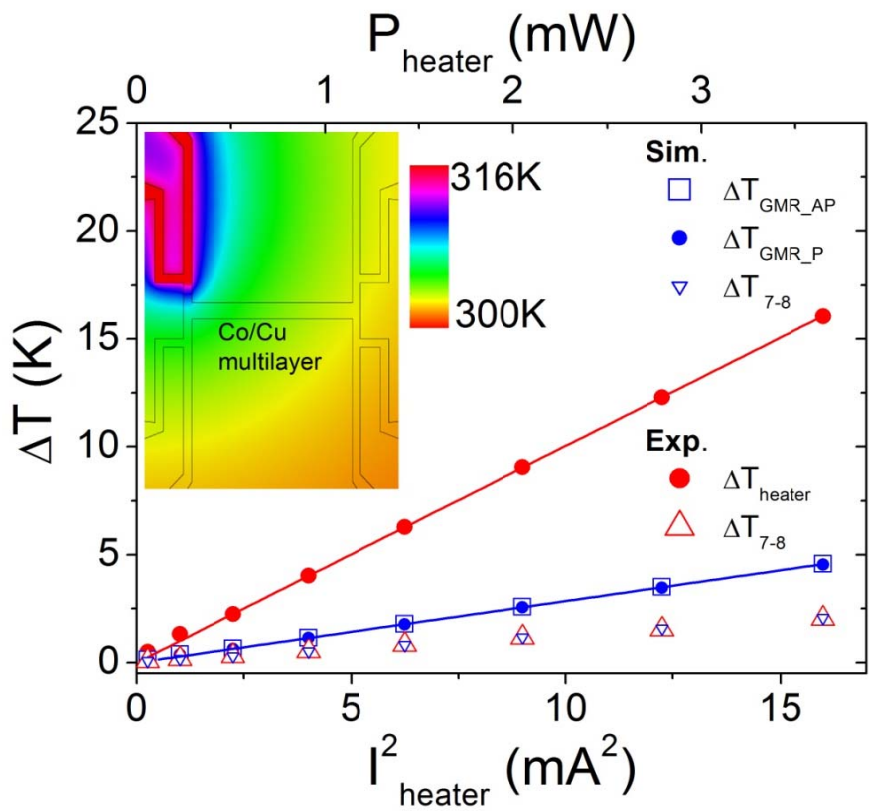

**Fig.3**

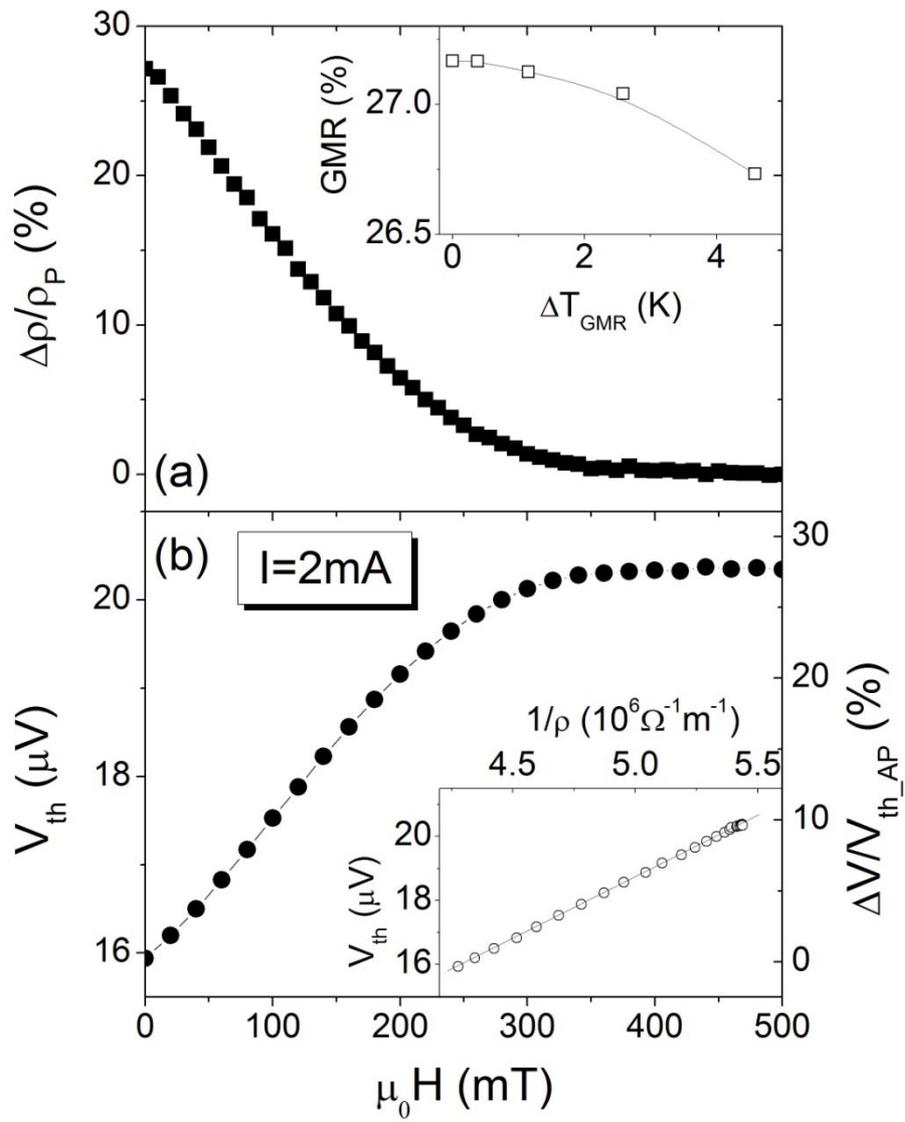

**Fig.4**

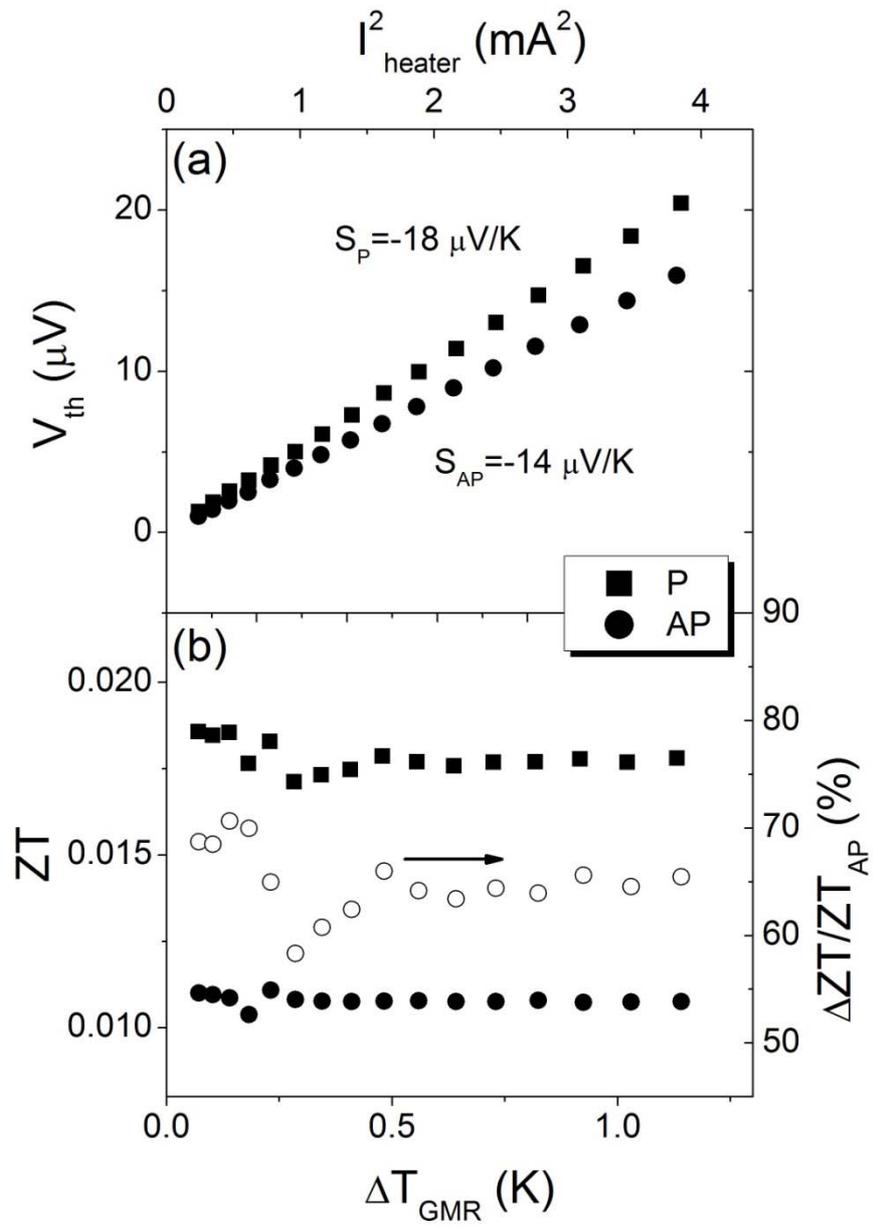